\documentclass[12pt]{article}
\usepackage{mathrsfs}
\usepackage{latexsym}
\usepackage{amsmath}
\usepackage{amsfonts}
\usepackage{amssymb}
\usepackage{amsthm}
\usepackage{epsfig}
\usepackage{subfigure}
\usepackage{arydshln}
\usepackage{multirow}
\usepackage{hyperref}
 
\usepackage{natbib}
\usepackage{times}
\usepackage[usenames]{color}
\bibpunct{(}{)}{;}{a}{,}{,}

\newtheorem{example}{Example}

\newcommand{\piP}{\pi_P(\cdot|\mathcal{X}_n)}
\newcommand{\piPhi}{\pi_{\phi}(\cdot|\mathcal{X}_n)}
\newcommand{\xiPhi}{\xi\{\pi_{\phi}(\cdot|\mathcal{X}_n)\}}
\newcommand{\piPhiT}{\pi_{\phi_T}(\cdot|\mathcal{X}_n)}
\newcommand{\xiPhiT}{\xi\{\pi_{\phi_T}(\cdot|\mathcal{X}_n)\}}

\newcommand{\xiPhijbn}{\xi\{\pi_{\phi}(\cdot|\mathcal{X}_{j,b,n})\}}

\newcommand{\PBB}{P_{n}^{\scalebox{0.5}{\rm BB}}}

\newcommand{\piPjbnResc}{\pi_P(\cdot|\mathcal{X}_{j,b,n}^\ast)}
\newcommand{\piPhijbnResc}{\pi_{\phi}(\cdot|\mathcal{X}_{j,b,n}^\ast)}

\newcommand{\piPBlbb}{\tilde{\pi}_P^{\scalebox{0.5}{\rm BLBB}}(\cdot|\mathcal{X}_{j,b,n})}
\newcommand{\piPhiBlbb}{\tilde{\pi}_{\phi}^{\scalebox{0.5}{\rm BLBB}}(\cdot|\mathcal{X}_{j,b,n})}
\newcommand{\PBlbb}{P_{j,b,n}^{\scalebox{0.5}{\rm BLBB}}}
\newcommand{\xiPhiBlbb}{\xi\{\tilde{\pi}_{\phi}^{\scalebox{0.5}{\rm BLBB}}(\cdot|\mathcal{X}_{j,b,n})\}}

\newcommand{\piPhiTBlbb}{\tilde{\pi}_{\phi_T}^{\scalebox{0.5}{\rm BLBB}}(\cdot|\mathcal{X}_{j,b,n})}
\newcommand{\xiPhiTBlbb}{\xi\{\tilde{\pi}_{\phi_T}^{\scalebox{0.5}{\rm BLBB}}(\cdot|\mathcal{X}_{j,b,n})\}}

\newcommand{\piPSdbb}{\pi_P^{\scalebox{0.5}{\rm SDBB}}(\cdot|\mathcal{X}_{n})}
\newcommand{\piPhiSdbb}{\pi_{\phi}^{\scalebox{0.5}{\rm SDBB}}(\cdot|\mathcal{X}_{n})}
\newcommand{\PSdbb}{P_{b,n}^{\scalebox{0.5}{\rm SDBB}}}
\newcommand{\xiPhiSdbb}{\xi\{\pi_{\phi}^{\scalebox{0.5}{\rm SDBB}}(\cdot|\mathcal{X}_{n})\}}

\newcommand{\piPhiTSdbb}{\pi_{\phi_T}^{\scalebox{0.5}{\rm SDBB}}(\cdot|\mathcal{X}_{n})}
\newcommand{\xiPhiTSdbb}{\xi\{\pi_{\phi_T}^{\scalebox{0.5}{\rm SDBB}}(\cdot|\mathcal{X}_{n})\}}

\newcommand{\PDP}{P^{\scalebox{0.5}{\rm DP}}}
\newcommand{\PDPn}{P_{n}^{\scalebox{0.5}{\rm DP}}}
\newcommand{\PBLDP}{P_{j,b,n}^{\scalebox{0.5}{\rm BLDP}}}
\newcommand{\PSDDP}{P_{b,n}^{\scalebox{0.5}{\rm SDDP}}}

\usepackage[labelfont=bf]{caption} 
\usepackage[labelsep=period]{caption}

\title{\bf Bayesian Bootstraps for Massive Data}
\author{{\sc Andr\'es F. Barrientos, V\'ictor Pe\~na}\footnote{Andr\'es F. Barrientos is Postdoctoral Associate, Department of Statistical Science, Duke University, Durham,
NC 27708 (email: afb26@stat.duke.edu) and, V\'ictor Pe\~na is Assistant Professor, Department of Information Systems and Statistics, Baruch College, New York, NY 10010.}}
\begin{document}
\date{\today}

\maketitle

\begin{abstract}
    {\color{black}{In this article, we present data-subsetting algorithms that allow for the approximate and scalable implementation of the Bayesian bootstrap. They are analogous to two existing algorithms in the frequentist literature: the bag of little bootstraps \citep[][]{kleiner;talwalkar;sarkar;jordan;2014} and the subsampled double bootstrap \citep[][]{sengupta;volgushev;shao;2016}. Our algorithms have appealing theoretical and computational properties that are comparable to those of their frequentist counterparts. Additionally, we provide a strategy for performing lossless inference for a class of functionals of the Bayesian bootstrap, and briefly introduce extensions to the Dirichlet Process.}}
    
\end{abstract}

\section{Introduction}\label{Intro}

Massive datasets are increasingly common in statistical applications, mainly because current computational technologies are capable of efficiently recording and storing large datasets. As a consequence, there is an increasing demand for statistical methods that can analyze large volumes of data. Quite often, a single computer cannot store big datasets into its internal memory, and statistical analyses can only be performed in smaller subsets of the original data. In such cases, we must use algorithms that combine statistical outputs from subsets to approximate the results we would obtain if we analyzed the full dataset at once.

In the frequentist literature, two scalable adaptations of the bootstrap have been proposed: the bag of little bootstraps \citep[BLB;][]{kleiner;talwalkar;sarkar;jordan;2014} and the subsampled double bootstrap \citep[SDB;][]{sengupta;volgushev;shao;2016}. These adaptations are based on data-subsetting. The BLB proceeds by splitting the data into subsets, bootstrapping within each subset, and averaging the summaries of the ``little'' bootstraps to get a global assessment of an estimator. To resemble the  ``big'' bootstrap (i.e., the bootstrap based on the full dataset), the little bootstraps need to be rescaled. The rescaling is achieved by modifying the parameters of the corresponding multinomial distribution to avoid extra computational cost. The SDB is an alternative to the BLB. The SDB proceeds by drawing random subsets from the entire dataset, running a rescaled bootstrap of size one within each subset, computing the root function for each rescaled bootstrap, and finally computing a summary of the bootstrapped values to obtain a global assessment of the estimator. As shown in \cite{kleiner;talwalkar;sarkar;jordan;2014} and \cite{sengupta;volgushev;shao;2016}, the BLB and SDB are computationally efficient and provide adequate assessments of uncertainty.

In this paper, we develop data-subsetting methods for the Bayesian bootstrap (BB) that are analogous to the BLB and the SDB. We will refer to these adaptations as the bag of little Bayesian bootstraps (BLBB) and the subsampled double Bayesian bootstrap (SDBB), respectively. The BB is a nonparametric model for probability measures proposed by \cite{rubin;1981} as a Bayesian analogue to the bootstrap \citep{efron;1979}. As discussed in \citeauthor{lyddon2017generalized} (\citeyear{lyddon2017generalized}, page 1), the BB represents a useful modeling technique to perform \textit{general Bayes updating}, which ``is a way of updating prior belief distributions that does not need the construction of a global probability model.'' For massive datasets, the BB is an appealing procedure for two main reasons: (1) it can accommodate complex features that are inherent to big data and (2) its implementation does not rely on recursive sampling algorithms, which are usually computationally expensive (e.g., Markov Chain Monte Carlo methods). In addition to the BLBB and SDBB, we present a strategy for performing lossless inference for a class of functionals of the BB. As a natural extension, we generalize our data-subsetting strategies to Dirichlet Processes (DP).

Our methods complement the growing literature on scalable Bayesian methods.  \cite{matthew;chen;yu;wyle;2015,taddy2016nonparametric} advocate for the use of the BB in massive datasets and approximate the distribution of certain functionals through Taylor series expansions and empirical Bayes procedures. Unlike these approaches, our proposal is based on data-subsetting. Most data-subsetting procedures consist of three steps: (1) the dataset is divided into subsets; (2) for each subset, a rescaled posterior distribution (subposterior) resembling the full posterior (i.e., the posterior distribution obtained by conditioning on the whole dataset) is obtained; and (3) the subposteriors are combined to either approximate the full posterior or a posterior summary of interest (e.g., a posterior variance). Methods based on subposteriors can be classified into two categories: methods that rescale the prior 
\citep[see e.g.,][]{
wang;dunson;2013, 
neiswanger;wang;xing;2014, 
wang;guo;heller;dunson;2015,
scott;blocker;bonassi;chipman;george;mcculloch;2016} 
and methods that rescale the likelihood 
\citep[see e.g.][]{
minsker;srivastava;lin;dunson;2014;paper, 
srivastava;li;dunson;2015, 
srivastava;cevher;trandinh;dunson;2015,
li;srivastava;dunson;2017}. Particularly, there are some similarities between the method proposed in \cite{li;srivastava;dunson;2017} and the BLBB: both methods focus on summaries of the posterior (not on the full posterior itself) and use the same rule to combine them. The methodology proposed in \cite{li;srivastava;dunson;2017} applies to a large class of models; however, its theoretical guarantees are proved under assumptions (e.g., Assumption 3) that are not verifiable for the BB.

The article is organized as follows. In Section~\ref{BLBB_SDBB}, we introduce and define the BB. Then, we describe the BLBB, the SDBB, and a strategy to perform lossless inference. Section~\ref{BLBB_SDBB} ends with a discussion of extensions for the DP. In Section~\ref{simulations}, we illustrate the performance of our methods in simulation studies. In Section~\ref{OPM2011}, we apply the BLBB and SDBB to model U.S. federal employees' wages from a subset of the Office of Personnel Management's datafile. In Section~\ref{A2012}, we use our methods to model whether or not households in the American Community Survey are paying for a {fire\/hazard\/flood} insurance. The paper concludes with a discussion and directions for future work.

\section{Data-subsetting strategies for the Bayesian Bootstrap}\label{BLBB_SDBB}

Throughout, we assume that the observations $\mathcal{X}_n = \{X_1, X_2, \, ... \, , X_n\}$, $X_i\in\mathbb{R}^p$, are independent and identically distributed given a probability measure $P$, which represents the data-generating mechanism. The BB is a probability model for $P$ given the data $\mathcal{X}_n$ which admits the stochastic representation 
\begin{align}\label{BB}
\PBB(\cdot) = \sum_{i=1}^n W_i \delta_{X_i}(\cdot),
\end{align}
where $(W_1,\ldots,W_n) \sim {\rm Dirichlet}(1,\ldots,1)$. The BB defines a distribution on probability measures which bypasses the traditional prior to posterior update, and it is nonparametric in the sense that no assumptions are made on the data-generating mechanism $P$ beyond conditional independence of the data given $P$ \citep{taddy2016nonparametric}. Additionally, the stochastic representation above allows us to obtain draws from $\PBB$ directly without resorting to Markov Chain Monte Carlo methods. However, when $n$ is massive and the full dataset cannot be loaded into memory, the BB cannot be easily implemented and approximations are needed (such as the ones presented in this article).

Several papers investigate the theoretical properties and methodological uses of the BB. From a theoretical point of view, various authors have studied its first and second-order asymptotic properties \citep{lo;1987,weng;1989}, proposed extensions and variations  \citep{lo;1991,kim;lee;2003,ishwaran;james;zarepour;2009}, and provided distributional characterizations of functionals of $\PBB$ \citep{gasparini;1995,choudhuri;1998,cifarelli;melilli;2000}. The connections between the BB and other processes have also been an object of study. In relation to Dirichlet processes, the BB is a limiting case of the posterior distribution of a DP (as the concentration parameter of the DP goes to zero, it converges weakly to the BB), so we can interpret it as a ``noninformative'' limit of the DP \citep{lo;1987,muliere;secchi;1996,gasparini;1995}.

There are numerous applied and methodological contributions that use the BB as a building block. For example, it has been used in areas such as
censored data \citep{lo;1993a,james;1997},  
finite population \citep{lo;1988},
quantile regression \citep{hahn;1997}, 
quantile estimation \citep{meeden;1993},  
multivariate regression \citep{heckelei;mittelhammer;2003},
ROC curve estimation \citep{gu;ghosal;roy;2008},
predictive modeling \citep{clyde;lee;2001,fushiki;2010},
synthetic data \citep{dong;elliott;raghunathan;2014}, 
tree-based modeling  \citep{matthew;chen;yu;wyle;2015}, 
high-dimensional inverse covariance matrix estimation \citep{datta;ghosh;2014},
causal studies \citep{graham;mccoy;stephens;2016},
multiple imputation \citep{rubin;schenker;1986,siddique;belin;2008,zhou;elliott;raghunathan;2016}, among others.

In addition to the above-mentioned favorable properties, the BB also has some potentially unappealing features. For example, $\PBB$ is almost surely discrete, and its support is limited to the observed data $\mathcal{X}_n$. At this point, we would like to stress that the goal of the article is not to provide an extensive discussion of the advantages and shortcomings of the BB, but rather to provide methodological tools to implement the model when $n$ is massive.

Let $\piP$ be the posterior distribution of $P$ given $\mathcal{X}_n$, which has a stochastic representation as in Expression~\eqref{BB}. We assume that the goal is to make posterior inferences about a functional of $P$ denote by $\phi$. We use the notation $\piPhi$ for the posterior distribution of $\phi(P)$. In the following subsections, we introduce methods for approximating $\xiPhi$, where $\xi$ is a summary of interest (e.g., mean, variance, length of credible intervals).

\subsection{Bag of little Bayesian bootstraps}\label{BLBB}

The BLBB is an adaptation of the bag of little bootstraps proposed by \cite{kleiner;talwalkar;sarkar;jordan;2014}. This procedure provides an approximation of $\xiPhi$ when $n$ is large, and it comprises three steps: divide, rescale, and combine. In the first step, we randomly split the dataset into subsets of size $b$ such that each subset can be stored in the internal memory of the computer. We define these subsets by generating a random partition $\mathcal{I}_{1,b,n},\ldots,\mathcal{I}_{n/b,b,n}$ of the set $\{1,\ldots,n\}$, where $|\mathcal{I}_{j,b,n}|=b$.
For ease of exposition, we assume that $n$ is a multiple of $b$.
 In the second step, we compute a rescaled version of the posterior distribution of $\phi(P)$ for each dataset $\mathcal{X}_{j,b,n} = \{X_i\}_{i \in \mathcal{I}_{j,b,n}}$, $j=1,\ldots,n/b$. In the third step, we combine the summaries found with the rescaled posteriors to obtain an approximation of $\xiPhi$.

The purpose of rescaling is to define a posterior distribution that resembles the one we would obtain with the full dataset. Without rescaling,  the next step after partitioning the dataset would be to compute $\xiPhijbn$, $j=1,\ldots,n/b$, and then combine these summaries using, for example, the average; that is, to approximate $\xiPhi$ by  $b/n\sum_{j=1}^{n/b}\xiPhijbn$. This strategy could work if the $\xiPhijbn$ provided a reasonable approximation of $\xiPhi$.  Unfortunately, this is often not the case. For example, if $\xi$ is the variance of $\piPhi$, then we expect $\xiPhi < \xiPhijbn$, which in turn implies that $\xiPhi < b/n\sum_{j=1}^{n/b}\xiPhijbn$. For this reason, we rescale the posterior distributions within each subset by following a standard strategy employed in several data-spliting based procedures \citep[e.g.,][]{kleiner;talwalkar;sarkar;jordan;2014,minsker;srivastava;lin;dunson;2014;paper,srivastava;cevher;trandinh;dunson;2015}. Each subset $\mathcal{X}_{j,b,n}$ is replicated $n/b$ times such that the replicated dataset contains $n$ instead of $b$ data points. Then, we obtain the posterior distribution associated with the replicated dataset, which we refer to as the ``rescaled posterior distribution.''

For a functional $\phi(P)$, the rescaled posterior distribution of $\phi(P)$ given $\mathcal{X}_{j,b,n}$ is defined as $\piPhiBlbb = \piPhijbnResc$, where $\mathcal{X}_{j,b,n}^\ast$ denotes the dataset $\mathcal{X}_{j,b,n}$ replicated $n/b$ times, so we have a new sample of size $n$. Note that rescaling $\piPhiBlbb$ can be achieved by rescaling the posterior distribution of $P$ given $\mathcal{X}_{j,b,n}$. The rescaled posterior distribution of $P$ given $\mathcal{X}_{j,b,n}$ is denoted $\piPBlbb = \piPjbnResc$. In this case, the distribution $\piPBlbb$ also has a stochastic representation of the form
\begin{align}\label{processBLBB}
\PBlbb(\cdot) = \sum_{i \in \mathcal{I}_{j,b,n}^\ast} W_{i,j} \delta_{X_i}(\cdot) \overset{d}{=} \sum_{i \in \mathcal{I}_{j,b,n}} W_{i,j}^\ast \delta_{X_i}(\cdot),
\end{align}
where  $(W_{i,j})_{i \in \mathcal{I}_{j,b,n}^\ast} \sim {\rm Dirichlet}(1,\ldots,1)$, $(W_{i,j}^\ast)_{i \in \mathcal{I}_{j,b,n}} \sim {\rm Dirichlet}(n/b,\ldots,n/b)$, $\overset{d}{=}$ denotes equality in distribution,  and $\mathcal{I}_{j,b,n}^\ast$ denotes the subset $\mathcal{I}_{j,b,n}$ replicated $n/b$ times. The process in Expression~\eqref{processBLBB} belongs to the class of BB clones proposed in \cite{lo;1991}. With this stochastic representation, we have $\phi(\PBlbb)| \mathcal{X}_{j,b,n} \sim \piPhiBlbb$. Although we compute the rescaled posterior using a replicated dataset of size $n$, the computational cost of drawing from $\piPhiBlbb$ is the same as of a BB with sample size equal to $b$. Thus, we propose approximating $\xiPhi$ by $b/n\sum_{j=1}^{n/b} \xiPhiBlbb$.

We provide theoretical guarantees for approximating summaries of $\phi_T(\PBB, \mathbb{P}_n)=$ \break $\sqrt{n}(T(\PBB) - T(\mathbb{P}_n))$ by summaries of $\phi_T(\PBlbb, \mathbb{P}_{j,b,n})=\sqrt{n}(T(\PBlbb) - T(\mathbb{P}_{j,b,n}))$, where $T$ is a functional and, $\mathbb{P}_n$ and $\mathbb{P}_{j,b,n}$ are the empirical measures associated with $\mathcal{X}_n$ and $\mathcal{X}_{j,b,n}$, respectively. Since $E[\PBB|\mathcal{X}_n]=\mathbb{P}_n$ and $E[\PBlbb|\mathcal{X}_{j,b,n}]=\mathbb{P}_{j,b,n}$, we can think of $T(\mathbb{P}_n)$ and $T(\mathbb{P}_{j,b,n})$ as measures of central tendency for the distribution of $T(\PBB)|\mathcal{X}_n$ and $T(\PBlbb)|\mathcal{X}_{j,b,n}$, respectively. The functional $\phi_T$ can be used to quantify uncertainty (e.g., estimate the length of intervals and measures of dispersion) associated with the functional $T$. The study of the asymptotic properties of $\phi_T(\PBB, \mathbb{P}_n)$ and $\phi_T(\PBlbb, \mathbb{P}_{j,b,n})$ requires understanding the asymptotic behavior of the processes $\sqrt{n}(\PBB - \mathbb{P}_n)$ and $\sqrt{n}(\PBlbb - \mathbb{P}_{j,b,n})$. These processes
belong to the class of {\it weighted bootstrap empirical processes} studied in Section 3.6.2 of \cite{vanderbvaart;wellner;1996}. 
The connection with {\it weighted bootstrap empirical processes} and an assumption on the differentiability of $T$ allow us to prove that $b/n\sum_{j=1}^{n/b}\xiPhiTBlbb$ suitably approximates $\xiPhiT$, where $\piPhiTBlbb$ and $\piPhiT$ denote the distributions of $\phi_T(\PBlbb, \mathbb{P}_{j,b,n})|\mathcal{X}_{j,b,n}$ and  $\phi_T(\PBB, \mathbb{P}_n)|\mathcal{X}_{n}$, respectively. The proof of this result is similar to the proof of Theorem 1 in \cite{kleiner;talwalkar;sarkar;jordan;2014}, and it also relies on the assumption that $T$ is Hadamard differentiable at $P_0$ (the data generation mechanism) and the existence of a $P_0$-Donsker class. The specific statement of our result and its proof can be found in Theorem 1 in the supplementary material. We show that even if we use $s$ instead of  $n/b$ subsets ($s<n/b$), the average $s^{-1}\sum_{j=1}^{s} \xiPhiTBlbb$ can provide a reasonable approximation of $\xiPhiT$. Figure \ref{Algorithms} describes the resulting Monte Carlo algorithm for the BLBB.

\subsection{Subsampled double Bayesian bootstrap}\label{SDBB}

The SDBB is the Bayesian analogue to the subsampled double bootstrap for massive data proposed by \cite{sengupta;volgushev;shao;2016}, which also provides an approximation of $\xiPhi$. In \cite{sengupta;volgushev;shao;2016}, the authors claim that the SDB outperforms the BLB in some scenarios with limited time budget, especially when it is only possible to run $s<n/b$ little bootstraps. Therefore, we would expect the same phenomenon to occur with the BLBB and SDBB.

Theorem 1 in the supplementary material shows that $s^{-1}\sum_{j=1}^{s} \xiPhiTBlbb$ can provide a reasonable approximation of $\xiPhiT$. However, the BLBB could be outperformed by the SDBB because the little Bayesian bootstraps only consider a unique partition of the dataset and, if the computational budget is limited, only a fraction of the dataset contributes to the analysis. We refer to this fraction as \textit{sample coverage}, a term we borrow from \cite{sengupta;volgushev;shao;2016}.

\begin{figure}[!ht]
\begin{center}
\includegraphics[scale=0.63]{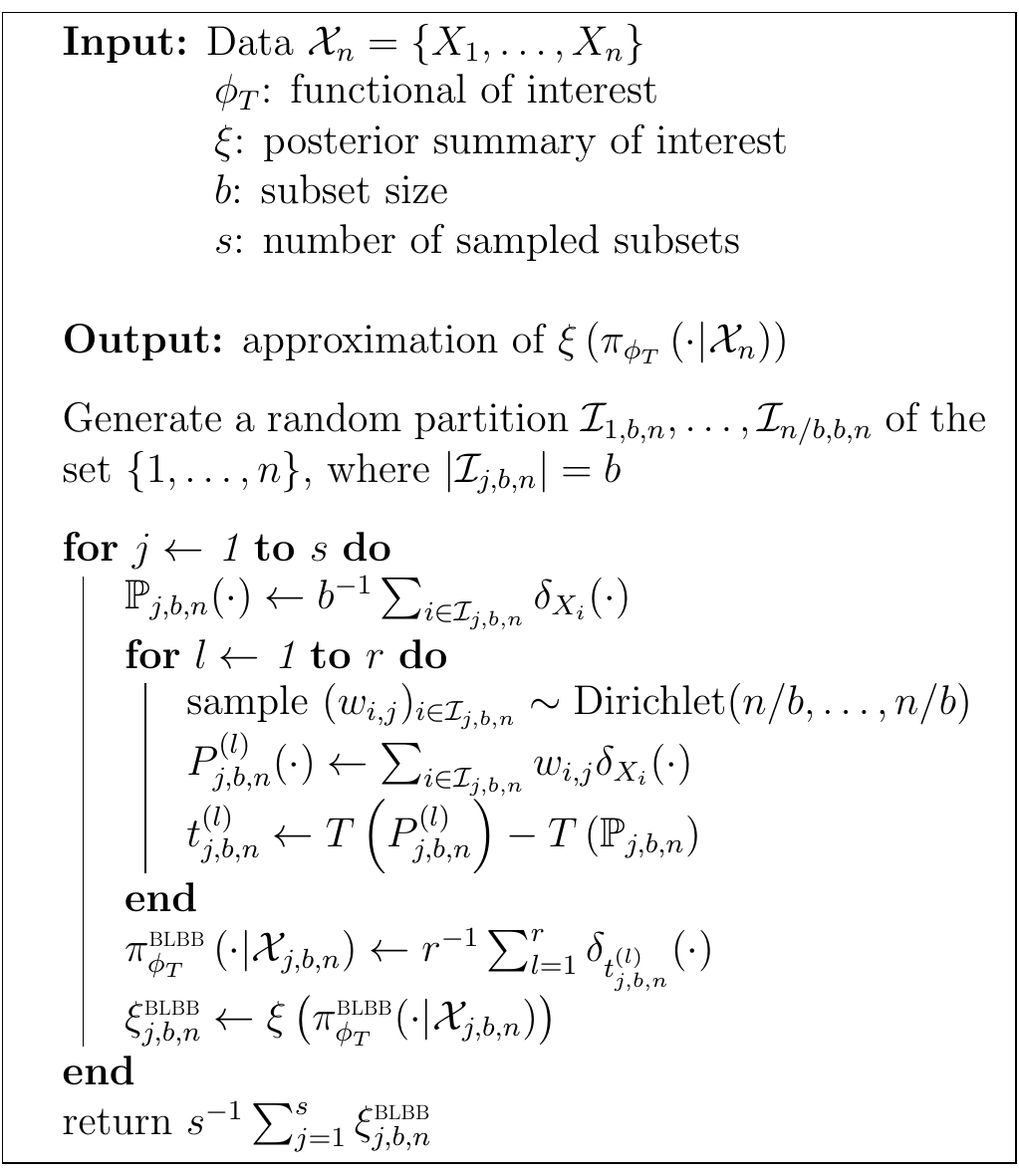}
\includegraphics[scale=0.63]{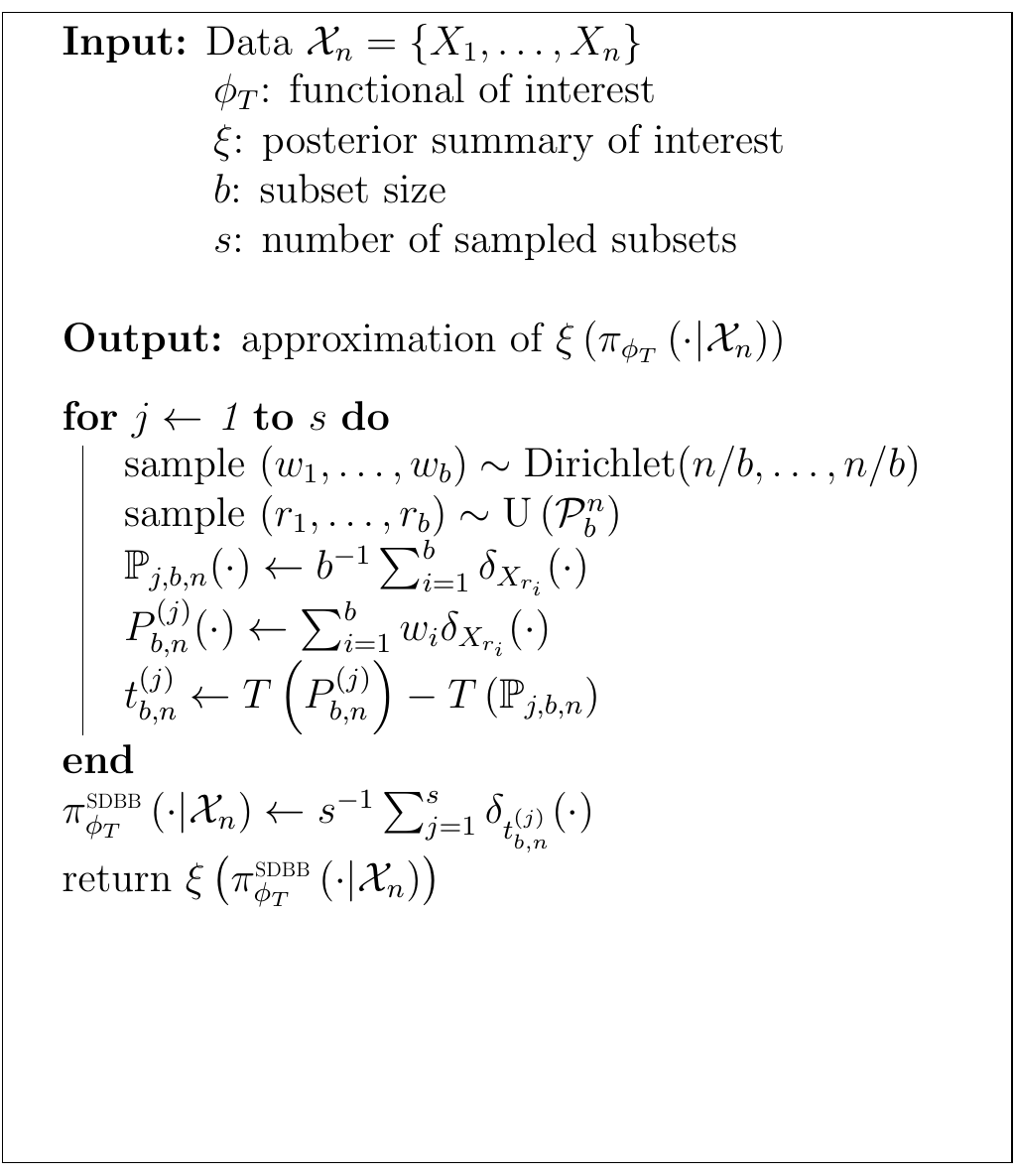}
\end{center}
\caption{\label{Algorithms} Monte Carlo algorithms for the BLBB (left) and SDBB (right).}
\end{figure}

The SDBB is a procedure that ensures a higher sample coverage compared to the BLBB and does not require using a partition of the dataset. Instead, this procedure uses random subsets of $\mathcal{X}_n$. Let $\mathcal{X}_{b,n} = \{X_{R_1},\ldots,X_{R_b}\}$ be representing the random subset, where $b\in\mathbb{N}$ is such that $\mathcal{X}_{b,n}$ can be stored in the internal memory of the computer, $R=(R_1,\ldots,R_b) \sim {\rm U}(\mathcal{P}_b^n)$, and ${\rm U}(\mathcal{P}_b^n)$ stands for the uniform distribution defined on the permutations of size $b$ of the elements $\{1,\ldots,n\}$. The SDBB runs a very little Bayesian bootstrap of size 1 for each drawn subset, so it has higher sample coverage than the BLBB. The use of subsets of $\mathcal{X}_n$ also requires a rescaling strategy, and we use the same one that was used for the BLBB. We approximate the posterior distribution of $P$ given $\mathcal{X}_{n}$ by $\piPSdbb$,  where $\piPSdbb$ is the distribution induced by the SDBB process. We define the SDBB process as
\begin{align*}
\PSdbb(\cdot) = \sum_{i \in \mathcal{I}_{b}^\ast} W_{i} \delta_{X_{R_i}}(\cdot) \overset{d}{=} \sum_{i=1}^b W_{i}^* \delta_{X_{R_i}}(\cdot)
\end{align*}
where $(W_{i})_{i \in \mathcal{I}_{b}^\ast} \sim {\rm Dirichlet}(1,\ldots,1)$, $(W_{i}^*)_{i=1}^b \sim {\rm Dirichlet}(n/b,\ldots,n/b)$, and $\mathcal{I}_{b}^\ast$ denotes the subset $\{1,\ldots,b\}$  replicated $n/b$ times. Let $\piPhiSdbb$ be the distribution of $\phi(\PSdbb)|\mathcal{X}_n$. Our proposal is to approximate $\xiPhi$ by  $\xiPhiSdbb$. We provide theoretical support for functionals of the form $\phi_T(\PBB, E[\PBB|\mathcal{X}_n])=\sqrt{n}(T(\PBB) - T(\mathbb{P}_n))$ (this is similar to the theoretical results in Section~\ref{BLBB}), where $T$ is a functional of interest. Theorem 2 in the supplementary material provides technical conditions under which $\xiPhiTSdbb$ approximates $\xiPhiT$, where $\piPhiTSdbb$ is the distribution of the functional $\phi_T(\PSdbb,  E[\PSdbb|\mathcal{X}_n,R])= \sqrt{n}(T(\PSdbb) - (E[\PSdbb|\mathcal{X}_n,R]))$ given $\mathcal{X}_n$, with $E[\PSdbb|\mathcal{X}_n,R]=b^{-1} \sum_{i=1}^b  \delta_{X_{R_i}}$. The technical conditions assumed for $T$  and $P_0$ in Theorem 2 are similar to those assumed for the BLBB. Theorem 2 is the counterpart of Theorem 1 in \cite{sengupta;volgushev;shao;2016}.  Figure \ref{Algorithms} describes the Monte Carlo algorithm for running the SDBB.

\subsection{Lossless inference for the Bayesian bootstrap}\label{lossless}

For a certain class of functionals, we can perform exact (lossless) inference for the BB after splitting the data. This strategy is based on decomposing the Dirichlet weights into Gamma random variables, and it is similar in spirit to the strategy devised for bagging in \cite{lee;clyde;2004}. The functionals for which lossless inference can be performed are of the form $\phi(\PBB) = g\left(\int\rho(x)\PBB(dx)\right)$ where $g$ is a measurable function, $\int\rho(x)\PBB(dx)=\left(\int\rho_1(x)\PBB(dx),\ldots,\right.$\break$\left.\int \rho_k(x)\PBB(dx)\right)$, $\rho = (\rho_1, \rho_2, \, \ldots \, , \rho_k)$, and $\rho_l$ is a function defined on the sample space. This class of functionals clearly contains moments and expectations of transformations, but it also contains other functionals such as weighted least squares estimators (as we show in the example below) or the instrumental variables estimator presented in Section 2 in \cite{chamberlain2003nonparametric} (see the supplementary material for further details).

\begin{example} (Weighted least squares) Let $Y_i \in \mathbb{R}$ be the outcome and $U_i \in \mathbb{R}^{p+1}$ be covariates, and assume that we want to model $E[Y_i|U_i]$ using a linear combination of the predictors. If the pairs are $(Y_i, U_i) | P \stackrel{\mathrm{iid}}{\sim} P$ and $P$ given the data is the BB, then we can define the least squares \textit{functional} (which, in this case, has a posterior distribution):
$$
\phi(\PBB)= \boldsymbol{\beta}^{\scalebox{0.5}{\rm BB}}_{{\rm lm},n}   = \underset{\boldsymbol{\beta}\in \mathbb{R}^{p+1}}{\mathrm{arg \, min}} \sum_{i=1}^n W_i(Y_i - U_i^\top \boldsymbol{\beta})^2.
$$
This functional has been used, for instance, in \cite{clyde;lee;2001} and \cite{taddy2016nonparametric}. We can rewrite
\begin{align*}
\boldsymbol{\beta}^{\scalebox{0.5}{\rm BB}}_{{\rm lm},n}  & = \left(\sum_{i=1}^n W_i U_i U_i^\top \right)^{-1}\left(\sum_{i=1}^n W_i U_{i} Y_i\right) \\
& =  \left(\int \rho_1(x) \PBB(dx)\right)^{-1}\left(\int \rho_2(x) \PBB(dx)\right)\\
& =  g\left(\int \rho(x) \PBB(dx)\right)
\end{align*}
where $x=(y,u)$, $y\in\mathbb{R}$, $u \in \mathbb{R}^{p+1}$, 
$\rho_1(x)=u u^\top$,
$\rho_2(x)=u y$,
$g(M,v) = M^{-1}v$,
$M$ is a $(p+1)\times (p+1)$-dimensional matrix, and $v$ is a $(p+1)$-dimensional vector. Thus, $\boldsymbol{\beta}^{\scalebox{0.5}{\rm BB}}_{{\rm lm},n}$ is included in the class of functionals for which lossless inference can be performed. 
\end{example}

The algorithm requires processing all the $n/b$ subsets, so it can be significantly slower than the BLBB or SDBB. A figure that summarizes the Monte Carlo algorithm for performing lossless inference can be found in the supplementary material.

\subsection{Extension to the Dirichlet Process} \label{DPforMD}

In the subsection, we extend the results in previous subsections for the Dirichlet process (DP), which we denote $X_i | P \overset{\mathrm{iid}}{\sim} P$, $i=1,\ldots,n,$ with $P  \sim {\rm DP}(\alpha, H)$. The hyperparameters of the DP are the base measure $H$ and the concentration parameter $\alpha > 0$. The base measure $H$ is the prior expectation of $P$, whereas $\alpha$ controls how concentrated $P$ is around $H$. The DP has the following explicit stochastic representation \citep{sethuraman;94}
 \begin{align*}
\PDP(\cdot) = \sum_{k=1}^\infty V_k \prod_{l=1}^{k-1} (1 - V_l) \, \delta_{A_k}(\cdot), \, \, \, \, \, V_k \overset{\mathrm{iid}}{\sim} \mathrm{Beta}(1,\alpha), \, \, \, \, \, A_k &\overset{\mathrm{iid}}{\sim} H,
\end{align*}
and it is a conjugate model:
\begin{equation*}
P  | \mathcal{X}_n \sim \mathrm{DP} \left(\alpha+n, \frac{\alpha}{\alpha+n} H + \frac{n}{\alpha+n} \mathbb{P}_n \right),
\end{equation*}
where $\mathbb{P}_n$ is the empirical probability measure. \cite{pitman;1996} shows that the posterior above can be represented as
\begin{equation} \label{posterior_DP}
\PDPn(\cdot) =  R_n\PDP(\cdot) + (1-R_n)  \PBB(\cdot), \, \, R_n \sim \mathrm{Beta}(\alpha,n). \\
\end{equation}
The random weight $R_n$, the prior measure $\PDP$ of the DP, and the BB $\PBB$ are independent given the data. For large $n$, the posterior measure of the DP is very close to the BB: if the same random $\PBB$ is to be used for an analysis with the BB and the DP,
\begin{align*}
\mathbb{P} [ d_{\mathrm{TV}}(\PBB, \PDPn) > \epsilon ] = \mathbb{P} [ R_n \, \sup_{A}| \PBB(A) - \PDPn(A)| > \epsilon] \le \mathbb{P}[R_n \ge \epsilon].
\end{align*}
For instance, if $0 < \epsilon < 1$ and $0 < \alpha \le 1$, the inequality above implies that $\mathbb{P} [ d_{\mathrm{TV}}(\PBB, \PDPn) > \epsilon ] \le (1-\epsilon)^{(n+1)}$.

From a more theoretical perspective, the representation in Equation~\eqref{posterior_DP} allows us to define an analogue of the BLBB and SDBB
\begin{align*}
\PBLDP  =  R_n\PDP(\cdot) + (1-R_n)\PBlbb(\cdot) \mbox{ and } 
\PSDDP  =  R_n\PDP(\cdot) + (1-R_n)\PSdbb(\cdot),
\end{align*}
where BLDP and SDDP stand for bag of little Dirichlet processes and subsampled double Dirichlet process, respectively, and $R_n \sim \mathrm{Beta}(\alpha,n)$. The Bernstein-von Mises results for the Dirichlet process that are available in the literature \citep[see e.g.,][]{lo1983weak,james;2008,varron;2014,castillo;nickl;2014} allow us to show that the asymptotic behavior of $\sqrt{n}(\PBLDP - E[\PBLDP|\mathcal{X}_{j,b,n}])$ and $\sqrt{n}(\PSDDP - E[\PSDDP|\mathcal{X}_n])$ is the same as that of $\sqrt{n}(\PDPn - E[\PDPn|\mathcal{X}_{n}])$, which is a parallel of the results found in the previous subsections (a proof can be found in the supplementary material). In order to have the same theoretical guarantees for a functional $\phi_T(\PBB,\mathbb{P}_n)$ (as described in previous subsections), the next step would be invoking the functional delta theorem. 
We finalize this subsection by noting that for the class of functionals that was defined in Subsection~\ref{lossless}, one can perform lossless inference for the Dirichlet-Multinomial process  \citep{kingman1975random,pitman;95}, which is an approximation to the Dirichlet Process \citep{ishwaran2002exact,muliere;secchi;1996}. The details can be found in the supplementary material.

\section{Numerical experiments} \label{simulations}

We compare the performance of the BLBB and SDBB in approximating posterior means and standard deviations, as well as the length of 95\% credible interval of functionals of the BB in linear, logistic, and mixed-effects regression. The sample sizes of the datasets are always equal to 10,000. We simulate data from the following models:
\begin{itemize}
    \item[-] \emph{Linear regression:} $Y_i = U_i^\top \boldsymbol{\beta}_0 + \epsilon_i$,
    \item[-] \emph{Logistic regression:} $Y_i \sim {\rm Bernoulli}(p_i),\  p_i=(1+\exp(-0.01U_i^\top \boldsymbol{\beta}_0+0.25))^{-1}$,
    \item[-] \emph{Mixed-effects regression:} $Y_{ij} = \alpha_j + U_{ij}^\top \boldsymbol{\beta}_0 + \epsilon_i, \ j=1,2,3$,
\end{itemize}
where  $i=1,\ldots,n$, $Y_i$ is the outcome, $U_i$ denotes a $(p+1)$-dimensional vector containing the predictors, and $\boldsymbol{\beta}_0 = (\beta_{0,0},\ldots,\beta_{p,0})^\top = (1, 1, \, ... \, , 1)$.   The first component of $U_i$  
is equal to 1 (acting as an intercept) and the remaining elements are independent and identically distributed as a Student-$t$ with 3 degrees of freedom. In the linear model, the errors $\epsilon_i$s are simulated from a Skew Normal distribution with location parameter $-0.71$, scale $1$, and slant $2$, which has mean 0 and is asymmetric. In the mixed model, we draw the random the effect $\alpha_j$ and error $\epsilon_i$ from a Skew Normal distribution with location parameter $-0.71$, scale $1$, and slant $2$ and from a Student-$t$ with 3 degrees of freedom, respectively. Finally, in linear regression, $p$ is taken to be 100, whereas in the case of logistic and mixed-effect regression $p$ is equal to 25. 

Throughout, we model the data as  $(Y_i, U_i) | P \stackrel{ \mathrm{iid} } {\sim} P$ and $P$ given the data using $\PBB$. 
We focus on the following 3 functionals that induce posterior distributions for the regression coefficients: 
\begin{itemize}
    \item[-] \emph{Linear regression:} the weighted least squares estimator (see, for example, \citealp{clyde;lee;2001} and \citealp{taddy2016nonparametric}),
    \begin{align} \label{funct:lm}
        \boldsymbol{\beta}^{\scalebox{0.5}{\rm BB}}_{{\rm lm},n}  
        & = \left(\sum_{i=1}^n W_i U_i U_i^\top\right)^{-1}\left(\sum_{i=1}^n W_i U_i Y\right),
    \end{align}
 
    \item[-] \emph{Logistic regression:} the robust estimator, \citep[see, for instance,][]{carroll1993robustness}, 
    \begin{align}\label{funct:lg}
    \boldsymbol{\beta}^{\scalebox{0.5}{\rm BB}}_{{\rm lg},n}  
    & = \underset{\boldsymbol{\beta}\in \mathbb{R}^{p}}{\mathrm{arg \, max}} \sum_{i=1}^n W_i \left[  
    -\log(1+\exp(U_i^\top \boldsymbol{\beta}))+Y_i U_i^\top \boldsymbol{\beta} \right], 
    \end{align}
    \item[-] \emph{Mixed-effects regression:} the weighted estimator,
    \begin{align}\label{funct:mx}
        \boldsymbol{\beta}^{\scalebox{0.5}{\rm BB}}_{{\rm mx},n}  
        = \left(\sum_{i=1}^n W_i {\bf U}_i^\top \hat{V} {\bf U}_i\right)^{-1}\left(\sum_{i=1}^n W_i {\bf U}_i^\top \hat{V} {\bf Y}_i\right),
    \end{align}
    where ${\bf U}_i = (U_{i,1},U_{i,2},U_{i,3})^\top$, ${\bf Y}_i = (Y_{i,1},Y_{i,2},Y_{i,3})^\top$,  and $\hat{V}$ is the MLE of the covariance matrix derived from the marginal likelihood of a mixed-effects model with random intercept and under Gaussian assumptions. \cite{welsh199713} and \cite{jacqmin2007robustness} discuss and assess the robustness of these type of functionals.
\end{itemize}

Now, we explain how the length of credible intervals, standard deviations, and mean for $\boldsymbol{\beta}^{\scalebox{0.5}{\rm BB}}_{{\rm lm},n} =
T(\PBB)$ are computed, using the notation introduced in Section~\ref{BLBB_SDBB} (the summaries for $\boldsymbol{\beta}^{\scalebox{0.5}{\rm BB}}_{{\rm lg},n}$ and $\boldsymbol{\beta}^{\scalebox{0.5}{\rm BB}}_{{\rm mx},n}$ are computed using the same procedure). In the simulation studies, 
$$\phi_T(\PBB, \mathbb{P}_n)=(T(\PBB)- T(\mathbb{P}_n))=(\boldsymbol{\beta}^{\scalebox{0.5}{\rm BB}}_{{\rm lm},n} - \hat{\boldsymbol{\beta}}_{{\rm lm},n}),$$
where $T(\PBB)=\boldsymbol{\beta}^{\scalebox{0.5}{\rm BB}}_{{\rm lm},n}$ and $T(\mathbb{P}_n)= \hat{\boldsymbol{\beta}}_{{\rm lm},n}$. 
We use the notation $\xi_{l,1}$, $\xi_{l,2}$, and $\xi_{l,3}$ for the 2.5th and 97.5th percentiles, and standard deviation of the $l$-th marginal distribution of a $(p+1)$-dimensional distribution. Thus, we define 
\begin{align}\nonumber
\xi_{l,k}^{\scalebox{0.5} {\rm BB}} &= \xi_{l,k}\{\piPhiT\}, \\\nonumber
\xi_{l,k}^{\scalebox{0.5} {\rm BLBB}} &= b/n\sum_{j=1}^{n/b}\xi_{l,k}\{\piPhiTBlbb\}, \\\nonumber
\xi_{l,k}^{\scalebox{0.5} {\rm SDBB}} &= \xi_{l,k}\{\piPhiTSdbb\},
\end{align}
where $\xi_{l,k}^{\scalebox{0.5} {\rm BLBB}}$ and $\xi_{l,k}^{\scalebox{0.5} {\rm SDBB}}$ are computed using the algorithms displayed in Figure \ref{Algorithms}.
With these summaries, we can compute the average relative errors of lengths of 95\% credible intervals and posterior standard deviations. For example, for the BLBB, we can compute the average relative error as
\begin{equation}\label{relative_error}
(p+1)^{-1} \sum_{l=0}^p \left|(\tilde{\xi}_{l,2}^{\scalebox{0.5} {\rm BLBB}}-\tilde{\xi}_{l,1}^{\scalebox{0.5}{\rm BLBB}})/
(\tilde{\xi}_{l,2}-\tilde{\xi}_{l,1})-1\right|,    
\end{equation}
and the same computation can be carried out for the SDBB by substituting $\tilde{\xi}_{l,1}^{\scalebox{0.5} {\rm BLBB}}$ and $\tilde{\xi}_{l,2}^{\scalebox{0.5} {\rm BLBB}}$ by $\tilde{\xi}_{l,1}^{\scalebox{0.5} {\rm SDBB}}$ and $\tilde{\xi}_{l,2}^{\scalebox{0.5} {\rm SDBB}}$. For the posterior mean, we define
\begin{align}\nonumber
\xi_{4}^{\scalebox{0.5} {\rm BB}} & = E[T(\PBB)|\mathcal{X}_n] = 
E[\boldsymbol{\beta}^{\scalebox{0.5}{\rm BB}}_{{\rm lm},n}|\mathcal{X}_n], \\\nonumber
\xi_{4}^{\scalebox{0.5} {\rm BLBB}} &= b/n\sum_{j=1}^{n/b}E[T(\PBlbb)|\mathcal{X}_{j,b,n}] = b/n\sum_{j=1}^{n/b}E[\boldsymbol{\beta}^{\scalebox{0.5}{\rm BLBB}}_{{\rm lm},n}|\mathcal{X}_{j,b,n}], \\\nonumber
\xi_{4}^{\scalebox{0.5} {\rm SDBB}} &= E[T(\PSdbb)|\mathcal{X}_n] = 
E[\boldsymbol{\beta}^{\scalebox{0.5}{\rm SDBB}}_{{\rm lm},n}|\mathcal{X}_n],
\end{align}
and quantify the bias by computing the absolute errors 
\begin{equation}\label{absolute_error}
\Vert \xi_{4}^{\scalebox{0.5} {\rm BB}} - \xi_{4}^{\scalebox{0.5} {\rm BLBB}}\Vert_1 \mbox{ and } \Vert \xi_{4}^{\scalebox{0.5} {\rm BB}} - \xi_{4}^{\scalebox{0.5} {\rm SDBB}}\Vert_1
\end{equation}
for the BLBB and SBBB, respectively. We use the errors in \eqref{relative_error} and \eqref{absolute_error} to assess the performance of the methods. We compare the results found after 1,000 samples from the BB (run on the full dataset), and the results shown are averages over 100 simulated datasets. In the case of the BLBB, the number of bootstrap samples within each subgroup is equal to $100$. On the other hand, the SDBB is run until 1,000 samples are drawn. For both methods, we set $b=n^\gamma$, $\gamma=0.6,\ 0.7,$ and $0.8$.

Given that the BLBB and SDBB have asymptotic guarantees (which are proved in Theorems 1 and 2 in the supplementary material), we compare the performance of the BLBB and SBBB with two procedures based on the asymptotic distributions of $T(\mathbb{P}_n)$ and $T(\mathbb{P}_{j,b,n})$. The first procedure uses an estimate of mean and variance of the asymptotic normal distribution of $T(\mathbb{P}_n)$ to compute the summaries of interest. We refer to this method based on asymptotic normality as AN. The second one relies on data-subsetting and, for each $j=1,\ldots,n/b$, uses an estimate of mean and variance (re-scaled by a factor of $b/n$) of the asymptotic normal distribution of $T(\mathbb{P}_{j,b,n})$ to compute an aggregated estimation of the corresponding summaries. We aggregate estimates using the average over $j=1,\ldots,n/b$. We refer to this method as ANS. The summaries computed under AN and ANS are compared to $\xi_{l,k}^{\scalebox{0.5} {\rm BB}}$, $k=1,2,3$, and $\xi_{4}^{\scalebox{0.5} {\rm BB}}$ using \eqref{relative_error}  and \eqref{absolute_error}. The estimated mean and variances of these asymptotic distributions as well as the functionals \eqref{funct:lm} and \eqref{funct:lg}, and the variance $\hat{V}$ in \eqref{funct:mx} are computed using the statistical software R \citep{R}. For linear and logistic regression, we used the function \texttt{lm} and  \texttt{glm} in the \texttt{stats} package; for mixed-effects regression, we used the function \texttt{lme} in library \texttt{nlme} \citep{nlme}. The simulations were performed on a computer with Intel(R) Xeon(R) CPU E5-2680 v3 @ 2.50GHz processor and 16 GB RAM and running R version 3.3.2.  We acknowledge that the results are contingent on our computing infrastructure and coding abilities. Section~\ref{MCsim} below explains the results of the simulation studies and Section~\ref{comp} contains a discussion on the computational overhead of the methods.

\subsection{Monte Carlo results} \label{MCsim}

In this subsection, we discuss and compare the Monte Carlo results obtained with the BB to those obtained with the BLBB, SDBB, ANS, and AN. Table \ref{tab:errors_simulations} displays the average relative and absolute errors associated with the summaries.

\begin{table}[!h]
\caption{\label{tab:errors_simulations} Average relative and absolute errors of approximate posterior summaries for the functionals in \eqref{funct:lm}-\eqref{funct:mx}. The average errors are computed over 100 simulated datasets and $(p+1)$ regression coefficients. Relative errors are reported for CIL and SD; absolute errors are reported for Mean.  CIL and SD stand for credible interval length and posterior standard deviation.}
\centering
\begin{tabular}{c|c|c|cccc}
Model            & Summary & $\gamma$  &  BLBB & SDBB & ANS  & AN  \\ \hline\hline
\multirow{9}{*}{Linear} & \multirow{3}{*}{CIL}  
& .6 & .043 & .088 & .375 & .046 \\
                        &   
& .7 & .045 & .062 & .140 & .046 \\
                        &   
& .8 & .053 & .048 & .073 & .046 \\ \cline{2-7}
                        & \multirow{3}{*}{SD}  
& .6 & .054 & .070 & .368 & .038 \\
                        &   
& .7 & .041 & .047 & .134 & .038 \\
                        &   
& .8 & .034 & .035 & .066 & .038 \\ \cline{2-7}
                        & \multirow{3}{*}{Mean}  
& .6 & .003 & .001 & .003 & $<$.001 \\
                        &   
& .7 & .001 & .001 & .001 & $<$.001 \\
                        &   
& .8 & .001 & .001 & .001 & $<$.001 \\ \hline\hline
\multirow{9}{*}{Mixed} & \multirow{3}{*}{CIL}  
& .6 & .089 & .107 & .120 & .063 \\
                        &   
& .7 & .069 & .093 & .104 & .063 \\
                        &   
& .8 & .064 & .056 & .091 & .063 \\ \cline{2-7}
                        & \multirow{3}{*}{SD}  
& .6 & .086 & .088 & .123 & .064 \\
                        &   
& .7 & .064 & .078 & .107 & .064 \\
                        &   
& .8 & .049 & .051 & .093 & .064 \\ \cline{2-7}
                        & \multirow{3}{*}{Mean}  
& .6 & .002 & .001 & .002 & $<$.001 \\
                        &   
& .7 & .002 & .001 & .002 & $<$.001 \\
                        &   
& .8 & .001 & .001 & .002 & $<$.001 \\ \hline\hline
\multirow{9}{*}{Logistic} & \multirow{3}{*}{CIL}  
& .6 & .130 & .208 & .254 & .022 \\
                        &   
& .7 & .037 & .093 & .102 & .022 \\
                        &   
& .8 & .033 & .050 & .050 & .022 \\ \cline{2-7}
                        & \multirow{3}{*}{SD}  
& .6 & .172 & .196 & .252 & .020 \\
                        &   
& .7 & .075 & .087 & .101 & .020 \\
                        &   
& .8 & .047 & .047 & .048 & .020 \\ \cline{2-7}
                        & \multirow{3}{*}{Mean}  
& .6 & .470 & .261 & .464 & .034 \\
                        &   
& .7 & .235 & .172 & .244 & .034 \\
                        &   
& .8 & .175 & .113 & .164 & .034\\ \hline\hline
\end{tabular}
\end{table}
 
In our simulations, ANS is outperformed by the BLBB and SDBB, but AN outperforms both of our methods. The good performance of AN is in agreement with our theoretical results which assert that, under appropriate regularity conditions, the posterior distribution of functionals of $\PBB$ will be approximately normal when the sample size is large enough. However, AN requires storing the full dataset in the internal memory, which is unfeasible in realistic applications with massive data. In other words, we propose to use the BLBB and SDBB when the data cannot be stored into the internal memory of a single computer, so AN is not a direct competitor.

As expected, Table \ref{tab:errors_simulations} shows that the performance of the proposed methods depends on the functional of interest. In linear and mixed-effects regression the errors are rather small, whereas in logistic regression the errors tend to be higher. For the functionals considered in this simulation study, we observe that the performance of the BLBB and SDBB suffers in scenarios where the functional of interest does not have a closed-form expression, such as logistic regression. Nonetheless, regardless of the functional, the BLBB and SDBB approximate the summaries of the BB better as the size of the subsets $b=n^\gamma$ increases, so we encourage users to use subsets as large as possible. Another component that can affect the quality of the approximation is the number of bootstrap samples used to compute the summaries. We recommend setting this number as large as possible.

While the bias of the SDBB is smaller than the bias of the BLBB in approximating posterior means, the BLBB is better than the SDBB in approximating credible interval lengths and posterior standard deviations. The bias of the BLBB can be attributed to the fact that it only considers a single partition of the dataset, which can be avoided by averaging over several random subsets (so the SDBB avoids this issue). However, the use of several random subsets adds an additional source of randomness that leads to wider credible intervals and larger standard deviations. Fortunately, as pointed out above, these differences are less worrisome as the size of the subsets increases.

\subsection{Computational considerations} \label{comp}

\begin{figure}[!h]
\begin{center}
\includegraphics[scale=0.315, page=1,trim={0,3cm 0 0,5cm 0},clip=true]{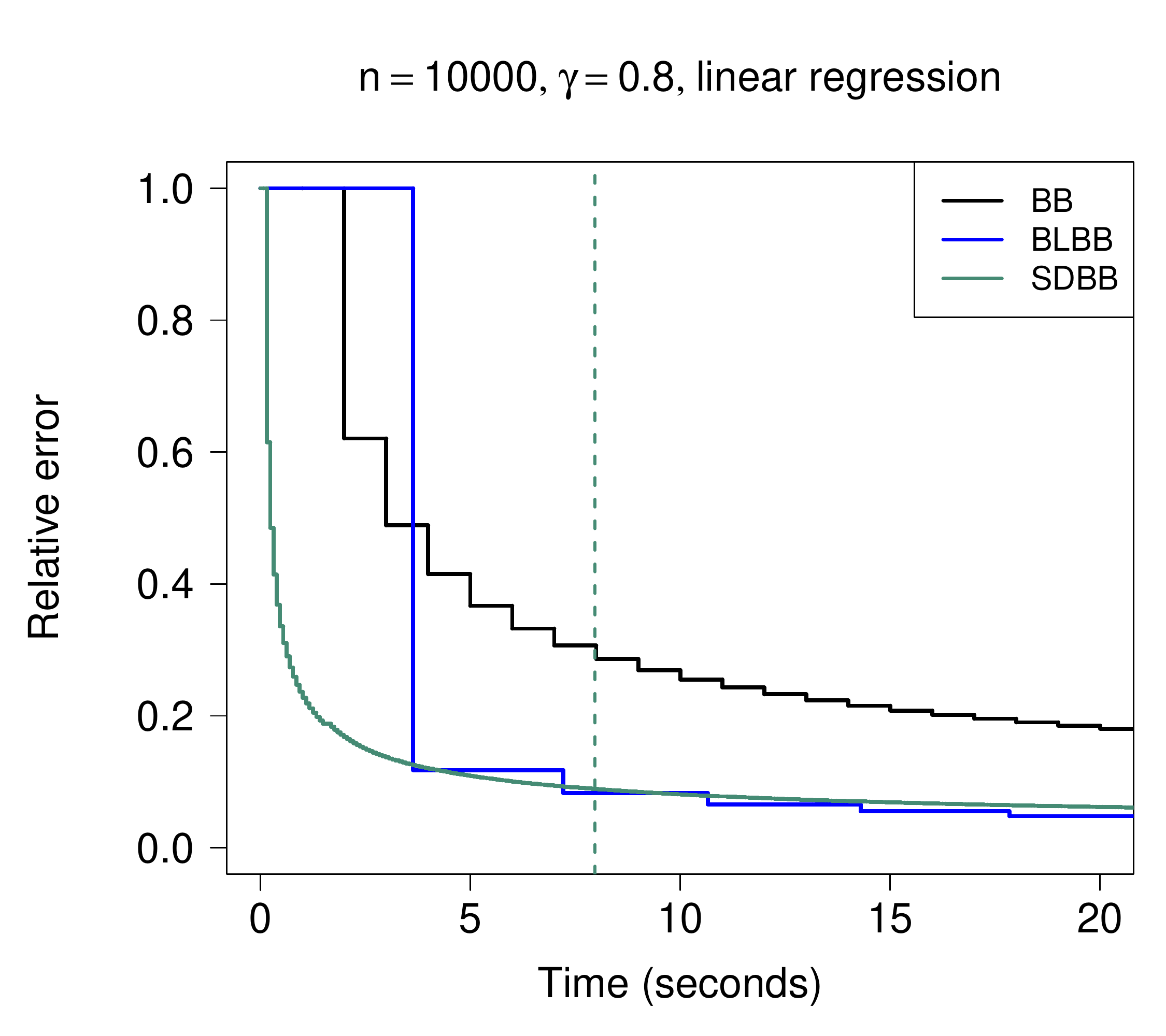}
\includegraphics[scale=0.315,page=1,trim={4,2cm 0 0,2cm 0},clip=true]{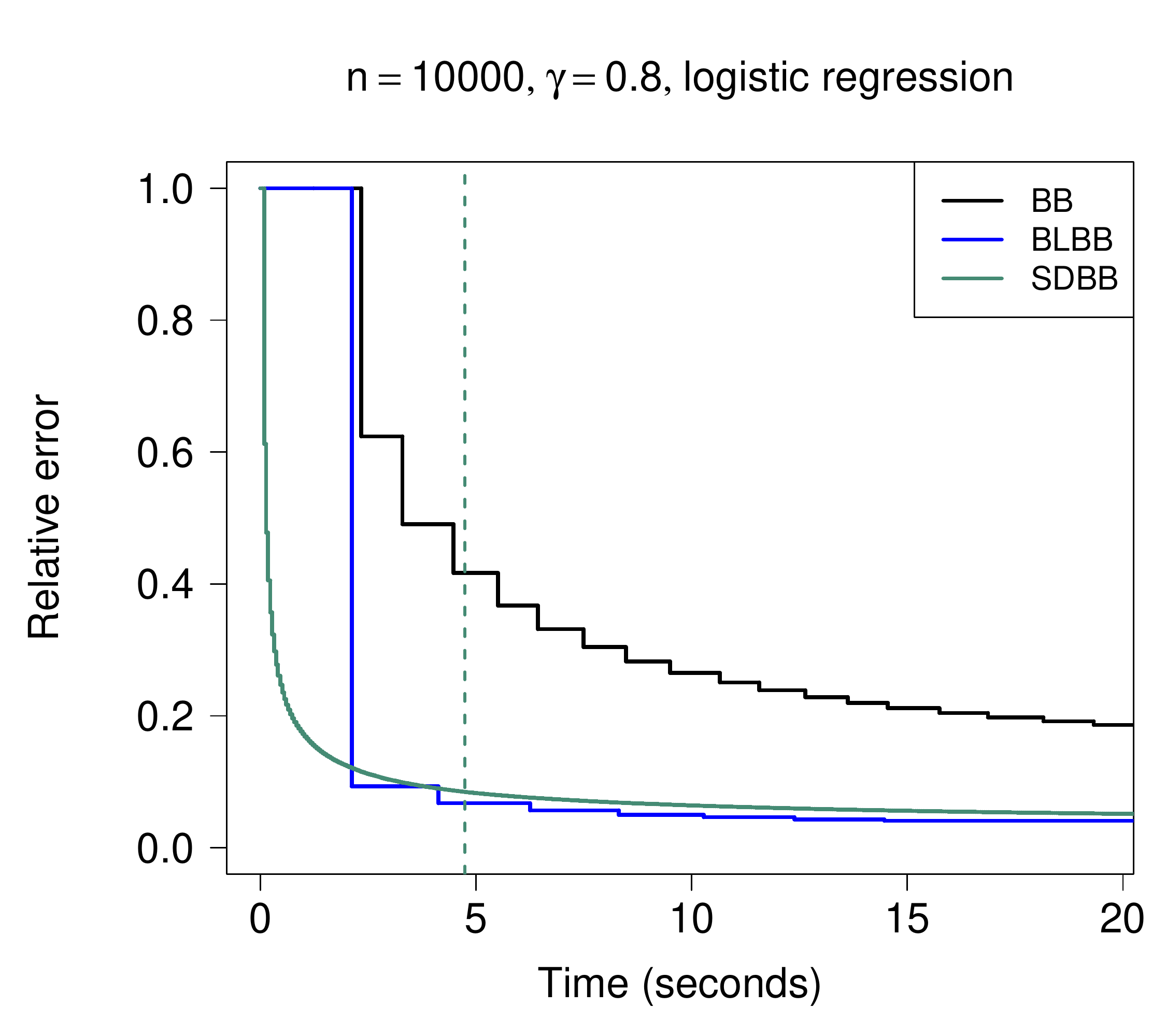}
\end{center}
\caption{\label{fig:relative_errors} 
Average relative errors and processing times associated with the length of credible intervals for $\boldsymbol{\beta}^{ \protect\scalebox{0.5}{\rm BB}}_{{\rm lm},n}$and $\boldsymbol{\beta}^{ \protect\scalebox{0.5}{\rm BB}}_{{\rm lg},n}$. The vertical green dashed line indicates when the SDBB has generated the 100th draw. The black line represents how fast the BB approximates the interval lengths based on 1,000 draws.}
\end{figure}

We compare the relative error of our methods with respect to 1,000 posterior draws from the BB (run on the full dataset), and we average our results over 100 simulated datasets. In the case of the BLBB, the number of bootstrap samples within each subgroup equals $100$, and the algorithm is run for 20 seconds (if all the subgroups are processed before 20 seconds, the algorithm stops). On the other hand, the SDBB runs for 20 seconds (unless 1,000 samples are sampled before 20 seconds, in which case the algorithm stops). For both methods, the subsets are processed sequentially using a single core. Figure \ref{fig:relative_errors} displays the results found for linear and logistic regression and credible interval lengths using $\gamma=0.8$. Due to the similarity of conclusions, we do not present the results for mixed-effects regression and other values of $\gamma$.

Figure \ref{fig:relative_errors} shows that the SDBB provides faster outputs than the BLBB if we do not wait until having 100 simulations to compute summaries of the posterior distribution. If we wait, the BLBB is faster than the SDBB. This phenomenon occurs because each iteration of the SDBB requires computing the functional twice, whereas the BLBB only computes the functional once. If we wait, the SDBB produces outputs just after the BLBB has processed the second subset. We recommend waiting until having some minimum number of samples because, otherwise, the estimates might not be reliable (in the sense that they will have high variance). Indeed, the figure shows that while the SDBB can produce outputs very quickly, reporting inferences with very few samples can have high relative error.

The computational cost of loading subsets of a massive dataset into memory is not always negligible and can play a crucial role in the performance of the proposed methods, particularly for the SDBB.  If $b$ is large enough, each subset takes a considerable amount of time to be loaded into memory, which can make the SDBB slower than the BLBB. More precisely, if we consider the same settings used in the simulation studies, the BLBB only has to load $n/b$ subsets, while the SDBB needs to load 1,000 subsets into memory. Hence, the use of the SDBB is limited to scenarios where, according to the available computational resources, the value of $b$ is not too large or the data transfer rate is not too low.

Although the BLBB and SDBB do not start reporting results at the same time (which depends on the size of $b$ and whether or not we impose a condition of having at least 100 draws from the SDBB), both provide results that stabilize rather quickly: in Figure~\ref{fig:relative_errors}, the relative errors tend to stabilize before the $n/b$ subsets for the BLBB and the 1,000 subsets for the SDBB have been processed. This is appealing in scenarios with limited access to computational resources. Even if the user can load the full dataset into memory, our methods will stabilize faster and might output better results than the BB. ANS is a valid alternative in this context, but users have to be willing to accept higher relative errors.

We conclude this section discussing some computational aspects related to the lossless procedure presented in Section~\ref{lossless}. While it is possible to perform lossless inference for $\boldsymbol{\beta}^{\scalebox{0.5}{\rm BB}}_{{\rm lm},n}$, it is significantly slower than the BLBB and SDBB. In our simulations, the BLBB and SDBB can produce excellent approximations in 20 seconds. Table~\ref{tab:lossless} shows computing times for the lossless method with different subgroup sizes and number of bootstrap samples that are drawn at a time from the subgroups (referred to as ``batch'' in the table). We show the average time spent sampling from the subsets (processing) and the time it takes to combine the results once all the samples are drawn (combining). The last column is an estimate of how long the procedure would take to generate 1,000 samples using the lossless method. We observe that smaller subgroups and bigger batches are preferable, but in any case the time it takes to generate 1,000 samples is significantly larger than the time that it takes to produce them using the BLBB and SDBB (and recall that, in this context, the BLBB and SDBB provide very good approximations, so the benefit of using a lossless method is minimal).

\begin{table}
\caption{Average computation times (in seconds) for lossless inference for $\boldsymbol{\beta}^{ \protect\scalebox{0.5}{\rm BB}}_{{\rm lm},n}$ and using different subgroup and batch sizes.}
\centering
\begin{tabular}{c|c|cccc }
Size of subgroup            & Batch  &  Processing & Combining & Full BB \\ \hline\hline
\multirow{2}{*}{$n^{0.6}$}  & 10  &  6 & 0.02 & 569 \\
                            & 100 & 96 & 0.46 & 966 \\ \hline
\multirow{2}{*}{$n^{0.7}$}  & 10  &  22 & 0.02 & 2203 \\
                            & 100 & 152 & 0.22 & 1526 \\ \hline
\multirow{2}{*}{$n^{0.8}$}  & 10  &  38 & 0.01 & 3865 \\
                            & 100 & 385 & 0.18 & 3854 \\\hline\hline
\end{tabular}
\label{tab:lossless}
\end{table}

\section{Applications with real-life datasets} \label{Applications}

In this section, we apply the BLBB and SDBB to 2 different datasets. The first one is the Office of Personnel Management's Central Personnel Data File (\url{https://www.opm.gov/}), which we refer to as the OPM dataset from now on. The OPM dataset is confidential and housed by the Protected Research Data Network at Duke University \url{https://oit.duke.edu/what-we-do/services/protected-network}. We consider two subsets of the data: i) a subset comprising employees that worked during 2011 and is referred to as OPM-2011, and ii) a subset comprising employees that worked without any interruption during ten years starting in 2002, which we refer to as OPM-10Y.  The second dataset includes public microdata files from the American Community Survey 2012 (ACS-2012), which can be downloaded from the United States Bureau of the Census (\url{http://www2.census.gov/acs2012_1yr/pums/}).

We study the performance of the BLBB and SDBB in approximating functionals of the posterior distribution of regression coefficients. We compare their 95\% credible intervals, posterior standard deviations, and posterior means (as we did in Section \ref{simulations}).  We approximate 95\% credible intervals using $\xi_{l,4}^{\scalebox{0.5} {\rm BLBB}} \pm (\xi_{l,2}^{\scalebox{0.5} {\rm BLBB}}-\xi_{l,1}^{\scalebox{0.5} {\rm BLBB}})/2$ and $\xi_{l,4}^{\scalebox{0.5} {\rm SDBB}} \pm (\xi_{l,2}^{\scalebox{0.5} {\rm SDBB}}-\xi_{l,1}^{\scalebox{0.5} {\rm SDBB}})/2$.

We also compare the BLBB and SDBB to the asymptotic methods ANS and AN, which were defined in Section~\ref{simulations}. For the OPM-2011 dataset, we focus on estimating coefficients from linear regression, i.e., $\boldsymbol{\beta}^{\scalebox{0.5}{\rm BB}}_{{\rm lm},n}$ (an analogous comparison for the $\tau$-quantile regression estimator proposed in \cite{chamberlain2003nonparametric} can be found in the supplementary material). For the OPM-10Y dataset, we use mixed-effects regression and aim to approximate summaries of the distribution of $\boldsymbol{\beta}^{\scalebox{0.5}{\rm BB}}_{{\rm mx},n}$. For the ACS-2012 dataset, we estimate coefficients of a logistic regression model, which we denote $\boldsymbol{\beta}^{\scalebox{0.5}{\rm BB}}_{{\rm lg},n}$.

\subsection{Office of Personnel Management} \label{OPM2011}

The OPM dataset comprises about 3.5 million employees and 29 variables recorded over 24 years. This dataset contains personnel records from employees that served in the federal U.\ S.\ government, including demographic variables (e.g., race, gender, and age) and relevant information related to their wages and careers (such as educational level).  Our response variable is the natural logarithm of the wages, and the predictors correspond to the variables whose effect is of interest (e.g., gender or race) along with other variables that are used to control for potential confounding factors (e.g., age and education level). In order to assess wage gaps between the levels of a feature of interest (e.g., between men and women or between races), researchers have run regression models and interpreted their coefficients {\citep[see][]{bolton;defigueiredo;2016a, bolton;defigueiredo;2016b, barrientos;etal;2018}}.  When those inferences cannot rely on parametric assumptions, uncertainty about the coefficients can be measured using nonparametric methods such as the BB.

For illustrative purposes, we use 2 random samples of 50,000 and 200,000 full-time employees from the OPM-2011 dataset, and 2 random samples of size 10,000 and 40,000 from the OPM-10Y dataset.  We include gender, race, educational level, and age as predictors. The levels for gender are male and female, whereas the levels for race are white, American indian/Alaskan native, Asian or pacific islander, black, and hispanic. Age and educational level are treated as numerical variables, and we include both linear and quadratic effects on the age of the employees. The datasets contain $22$ different educational categories. For ease of interpretation, we collapse the categories into a single continuous measure of the years of education attained by an individual past high school. Race is treated as a categorical variable, and the baseline cannot be disclosed because the dataset is confidential. For the OPM-10Y, we add a predictor representing the year at which the observation was collected. The regression coefficient $\boldsymbol{\beta}^{\scalebox{0.5}{\rm BB}}_{{\rm mx},n}$ is computed using Expression \eqref{funct:mx} where $\hat{V}$ is computed using a mixed-effects model with a random intercept and slope for year, and assuming that each employee represents a level of the grouping factor.

We compare the results obtained with the BLBB, SDBB, ANS, and AN with those obtained after drawing 1,000 samples from the BB ran on the full dataset. The analysis was performed on a computer with a Intel(R) Xeon(R) CPU E5-2699 v8 @ 2.20GHz processor and 236 GB RAM (running R version 3.3.3). We consider $b = n^{\gamma}$ with $\gamma = 0.6$, $0.7$, and $0.8$.

\subsubsection{OPM-2011}

A table containing the relative and absolute errors associated with $\boldsymbol{\beta}^{\scalebox{0.5}{\rm BB}}_{{\rm lm},n}$ can be found in the supplementary material. The relative errors for interval lengths and standard deviations are rather small (less than 0.05) with all methods.  The biases (absolute errors) are also small (less than 0.015). In practice, the BLBB and ANS can give biased results, depending on the partition. In almost all of the cases, the BLBB and SDBB produced errors that are smaller or equal than the errors of ANS and AN.

In this application, the BLBB and SDBB provide satisfactory results for the values of $n$ and $\gamma$ we considered, and the effect of increasing $n$ and $\gamma$ is not particularly noticeable. On the other hand, the asymptotic methods improve their performance when $n$ and $\gamma$ are increased. Figure \ref{fig:CredibleIntervals} displays credible intervals for  coefficients associated with race. All the methods provide satisfactory approximations.

\subsubsection{OPM-10Y} \label{mixed}

Table \ref{tab:OPM_mx} contains the relative and absolute errors associated with $\boldsymbol{\beta}^{\scalebox{0.5}{\rm BB}}_{{\rm mx},n}$. The errors are rather large for $n=$10,000 and $\gamma=0.6$. However, as $n$ or $\gamma$ increase, the BLBB and SDBB are better approximations of the BB. For $n=$40,000 and $\gamma=0.8$, the errors are fairly small. For this specific dataset, none of the methods outperforms the others, but we observe some patterns. For example, for $\gamma=0.6$, the BLBB assesses uncertainty better than the SDBB, and for $n$=10,000, the bias associated with SDBB is smaller than the bias of the BLBB.

The relative errors for interval lengths and standard deviations are quite large with the asymptotic methods if we use the asymptotic estimator of the variance proposed in  \cite{jacqmin2007robustness}.  We also consider the sandwich estimator proposed in \cite{liang1986longitudinal}, which is implemented in the R function \texttt{vcovCR.lme} of the library \texttt{clubSandwich} \citep{clubSandwich}.The results with the sandwich estimator are labeled ANS-Sand and AN-Sand. Table  \ref{tab:OPM_mx} also contains the relative and absolute errors associated with these asymptotic methods. The errors associated with ANS-Sand are similar to the errors found with the BLBB. 
 
\begin{figure}[!h]
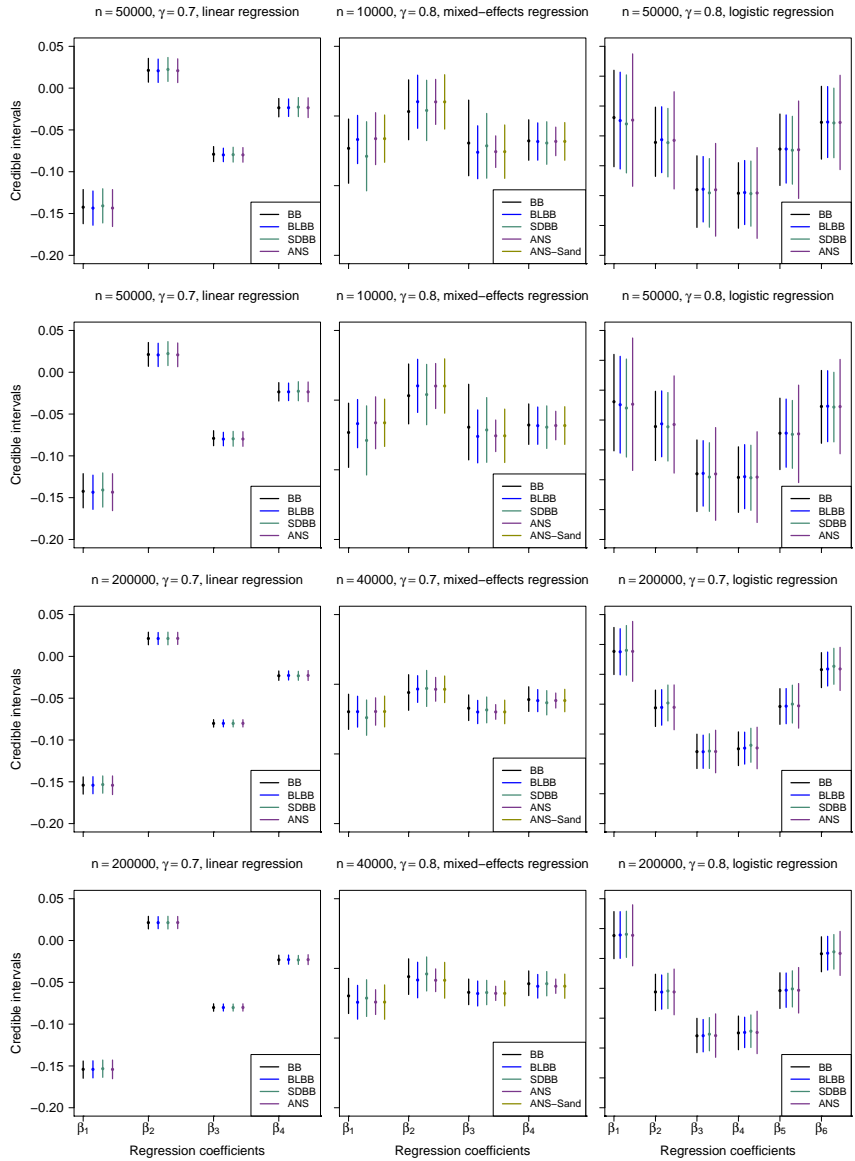

\begin{center}
\includegraphics[scale=0.215,page=2,trim={0 2,9cm 0,5cm 0},clip=true]
{Graphs/lm_n50000.pdf}
\includegraphics[scale=0.215,page=3,trim={3,9cm 2,9cm 0,5cm 0},clip=true]
{Graphs/mixed_n100000.pdf}
\includegraphics[scale=0.215,page=3,trim={3,9cm 2,9cm 0,5cm 0},clip=true]
{Graphs/Logistic_n50000.pdf}

\includegraphics[scale=0.215,page=2,trim={0 2,9cm 0,5cm 0},clip=true]
{Graphs/lm_n50000.pdf}
\includegraphics[scale=0.215,page=3,trim={3,9cm 2,9cm 0,5cm 0},clip=true]
{Graphs/mixed_n100000.pdf}
\includegraphics[scale=0.215,page=3,trim={3,9cm 2,9cm 0,5cm 0},clip=true]
{Graphs/Logistic_n50000.pdf}

\includegraphics[scale=0.215,page=2,trim={0 2,9cm 0,5cm 0},clip=true]
{Graphs/lm_n200000.pdf}
\includegraphics[scale=0.215,page=2,trim={3,9cm 2,9cm 0,5cm 0},clip=true]
{Graphs/mixed_n400000.pdf}
\includegraphics[scale=0.215,page=2,trim={3,9cm 2,9cm 0,5cm 0},clip=true]
{Graphs/Logistic_n200000.pdf}

\includegraphics[scale=0.215,page=2,trim={0 0,5cm 0,5cm 0},clip=true]
{Graphs/lm_n200000.pdf}
\includegraphics[scale=0.215,page=3,trim={3,9cm 0,5cm 0,5cm 0},clip=true]
{Graphs/mixed_n400000.pdf}
\includegraphics[scale=0.215,page=3,trim={3,9cm 0,5cm 0,5cm 0},clip=true]
{Graphs/Logistic_n200000.pdf}
\end{center}
\caption{\label{fig:CredibleIntervals}  Credible intervals for regression coefficients, two different values of $n$, and $\gamma \in \{0.7,0.8\}$. The first and second columns display intervals associated with the levels of the predictor race from the OPM-2011 and OPM-10Y datasets, respectively.  The third column displays intervals associated with the levels of the predictor indicating how long the families lived in the household, ACS-2012 dataset. }
\end{figure}

Figure \ref{fig:CredibleIntervals} displays credible intervals for coefficients associated with different levels of race. Again, we observe that increasing $n$ or $\gamma$ improves the quality of the approximation. For $n=$40,000, we do not observe big discrepancies between the credible intervals based on the BB and the BLBB, SDBB, or ANS-Sand; however, the intervals with ANS tend to be too narrow. Additionally, in Figure  \ref{fig:CredibleIntervals}  we observe that the most problematic coefficient is the first one (represented as $\beta_1$ in the figure), which is associated with a category with a small observed frequency (2\%). If a variable has some levels that are highly infrequent, approximating their coefficients via subsetting can be problematic: for example, some of the subsets might have extremely small frequencies for some levels, and some levels might not be observed at all. 
    
In this specific context (OPM dataset, linear and mixed-effect regressions), our general recommendation to take subset sizes as large as possible leads to reasonable performance of the BLBB and the SDBB. 

\begin{table}
\caption{Average relative and absolute errors of approximate posterior summaries for $\boldsymbol{\beta}^{ \protect\scalebox{0.5}{\rm BB}}_{{\rm mx},n}$, OMP-10Y dataset. The average errors are computed over all regression coefficients. Relative errors are reported for CIL and SD; absolute errors are reported for Mean. CIL and SD stand for credible interval length and posterior standard deviation.}\label{tab:OPM_mx}
\centering
\begin{tabular}{c|c|c|cccccc }
n            & Summary & $\gamma$  &  BLBB & SDBB & ANS  & AN & ANS-Sand  & AN-Sand  \\ \hline\hline
\multirow{9}{*}{10,000} & \multirow{3}{*}{CIL}  
& .6 & .190 & .307 & .365 & .259 & .194 & .009 \\
                        &   
& .7 & .107 & .128 & .216 & .259 & .100 & .009 \\
                        &   
& .8 & .095 & .044 & .240 & .259 & .088 & .009 \\ \cline{2-9}
                        & \multirow{3}{*}{SD}  
& .6 & .197 & .249 & .370 & .259 & .198 & .011 \\
                        &   
& .7 & .100 & .118 & .218 & .259 & .100 & .011 \\
                        &   
& .8 & .087 & .033 & .240 & .259 & .084 & .011 \\ \cline{2-9}
                        & \multirow{3}{*}{Mean}  
& .6 & .204 & .162 & .204 & .002 & .204 & .002 \\
                        &   
& .7 & .055 & .053 & .054 & .002 & .054 & .002 \\
                        &   
& .8 & .024 & .012 & .023 & .002 & .023 &  .002 \\ \hline\hline
\multirow{9}{*}{40,000} & \multirow{3}{*}{CIL}  
& .6 & .077 & .105 & .197 & .269 & .075 & .009 \\
                        &   
& .7 & .076 & .024 & .242 & .269 & .072 & .009 \\
                        &   
& .8 & .022 & .027 & .259 & .269 & .019 & .009 \\ \cline{2-9}
                        & \multirow{3}{*}{SD}  
& .6 & .067 & .079 & .199 & .267 & .068 & .007 \\
                        &   
& .7 & .067 & .036 & .239 & .267 & .064 & .007 \\
                        &   
& .8 & .010 & .014 & .257 & .267 & .011 & .007 \\ \cline{2-9}
                        & \multirow{3}{*}{Mean}  
& .6 & .055 & .062 & .054 & $<$0.001 & .054 & $<$0.001 \\
                        &   
& .7 & .008 & .013 & .008 & $<$0.001 & .008 & $<$0.001 \\
                        &   
& .8 & .011 & .004 & .011 & $<$0.001 & .011 & $<$0.001 \\  \hline\hline
\end{tabular}
\end{table}

\subsection{American Community Survey} \label{A2012}

The ACS-2012 dataset contains records of {1,477,091} households in the United States collected in 2012. The data were collected with the goal of making inferences about population demographics and housing characteristics. In this application, we only use the information at the household level and not at the individual level, and we model a binary outcome (response variable) that indicates whether or not the household is paying for a {fire/hazard/flood} insurance. We regress our binary outcome on a set of numerical and categorical predictors. As numerical predictors, we consider number of people living in the household, number of bedrooms, number of rooms, number of vehicles, and household income (in the past 12 months). The categorical predictors are lot size, yearly food stamp/Supplemental Nutrition Assistance Program (SNAP) recipiency, house heating fuel, presence and age of children, and a discretized variable that indicates how long the families lived in the household. This set of predictors leads to a regression with 21 coefficients.

We use two subsets of the ACS-2012 dataset of $n=$50,000 and 200,000 complete cases records that correspond to households that are located in the Northeast of the United States and are not rentals. We estimate the posterior mean and standard deviation, quantiles (2.5\% and 97.5\%), and lengths of the resulting 95\% credible intervals for  $\boldsymbol{\beta}^{\scalebox{0.5}{\rm BB}}_{{\rm lg},n}$. We compare the results obtained with the BLBB, SDBB, and the asymptotic methods ANS and AN with those obtained from running 1,000 samples from the BB ran on the full dataset.

Table \ref{tab:ACS_lg} contains the relative and absolute errors associated with $\boldsymbol{\beta}^{\scalebox{0.5}{\rm BB}}_{{\rm lg},n}$. The errors are moderately small ($<$0.09) even for $n=$10,000 and $\gamma=0.6$. We also observe that, as either $n$ or $\gamma$ increase, the BLBB and SDBB provide better results. In this application, we cannot conclude that the BLBB or SDBB uniformly outperforms the other; for $\gamma=0.6$, the BLBB is better at assessing uncertainty than the SDBB, whereas for $n$=50,000, the bias associated with SDBB is smaller than the bias of the BLBB. These patterns were also observed in the OPM-10Y dataset, where we used a mixed-effects model (see Section~\ref{mixed}). The relative errors for estimating interval lengths and standard deviations with the ANS and AN are large. The errors decrease as $n$ increases, but they are never smaller than the errors of the BLBB and SDBB. The bias (absolute error) associated with AN is very small, whereas the bias associated with ANS is similar to the bias of the BLBB. This is not surprising because we are using the same partition with both methods (BLBB and ANS).

Figure \ref{fig:CredibleIntervals} displays credible intervals for the coefficients of the variable that indicates how long the families lived in the household. We choose to show these intervals because the observed frequencies of some of the levels are low, so they are particularly hard to estimate with subsetting methods. ANS is more sensitive to this specific issue (which is even worse when $b$ is small) than the BLBB and SDBB. In general, we observe that ANS tends to output intervals that are too wide.

\begin{table}[!h]
\caption{Average relative and absolute errors of approximate posterior summaries for $\boldsymbol{\beta}^{ \protect\scalebox{0.5}{\rm BB}}_{{\rm lg},n}$, ACS-2012 dataset. The average errors are computed over all regression coefficients. Relative errors are reported for CIL and SD; absolute errors are reported for Mean. CIL and SD stand for credible interval length and posterior standard deviation.}\label{tab:ACS_lg}
\centering
\begin{tabular}{c|c|c|cccc }
n            & Summary & $\gamma$  &  BLBB & SDBB & ANS  & AN  \\ \hline\hline
\multirow{9}{*}{50,000} & \multirow{3}{*}{CIL}  
& .6 & .033 & .087 & .844 & .378 \\
                        &   
& .7 & .034 & .027 & .286 & .378 \\
                        &   
& .8 & .045 & .029 & .301 & .378 \\ \cline{2-7}
                        & \multirow{3}{*}{SD}  
& .6 & .056 & .082 & .850 & .374 \\
                        &   
& .7 & .021 & .026 & .282 & .374 \\
                        &   
& .8 & .026 & .024 & .297 & .374 \\ \cline{2-7}
                        & \multirow{3}{*}{Mean}  
& .6 & .491 & .414 & .480 & .010 \\
                        &   
& .7 & .191 & .168 & .173 & .010 \\
                        &   
& .8 & .080 & .069 & .060 & .010 \\ \hline\hline
\multirow{9}{*}{200,000} & \multirow{3}{*}{CIL}  
& .6 & .020 & .045 & .307 & .280 \\
                        &   
& .7 & .034 & .033 & .218 & .280 \\
                        &   
& .8 & .036 & .032 & .240 & .280 \\ \cline{2-7}
                        & \multirow{3}{*}{SD}  
& .6 & .029 & .041 & .298 & .271 \\
                        &   
& .7 & .018 & .023 & .210 & .271 \\
                        &   
& .8 & .022 & .027 & .232 & .271 \\ \cline{2-7}
                        & \multirow{3}{*}{Mean}  
& .6 & .089 & .118 & .087 & .009 \\
                        &   
& .7 & .024 & .063 & .024 & .009 \\
                        &   
& .8 & .016 & .021 & .015 & .009 \\ \hline\hline
\end{tabular}
\end{table}

\section{Discussion}

We have presented the BLBB and SDBB as two data-subsetting procedures to approximate the BB.  The BLBB and SDBB are analogous to the BLB \citep{kleiner;talwalkar;sarkar;jordan;2014} and SDB \citep{sengupta;volgushev;shao;2016}. The proposed procedures have theoretical and computational properties that are comparable to those of their frequentist counterparts. The performance of the methods has been illustrated and compared in simulation studies and real datasets. 
Although both the BLBB and SDBB are computationally efficient, the BLBB is preferable in scenarios where the computational cost of loading the subsets into memory is high. A similar conclusion can be drawn if the BLB and SDB are compared in an analogous setting. We observe that the BLBB approximates the uncertainty of the BB better than the SDBB, whereas the SDBB provides better approximations of point estimates than the BLBB. If the subsets can be loaded into memory reasonably fast and the functional of interest can be computed quickly, we recommend running rescaled bootstraps for some subsets to check if the posterior distributions of the functionals are similar. If they are, the SDBB will approximate the uncertainty as well as the BLBB does but with less bias; if they are not, the SDBB will overestimate uncertainty estimates and we recommend using the BLBB.

The performance of the methods depends on the size of the subsets and the functional of interest. In general, we observe that increasing subset sizes improves the approximation. This relationship between the quality of the approximation and subset size is not particular to our procedures; in fact, it is a common issue of data-subsetting methods. %
In addition to the BLBB and SDBB, we provide a strategy for performing lossless inference for functionals that can be expressed as functions of expectations with respect to the probability measure of the BB. This class is larger than one would expect at first glance: it includes, for instance, the weighted least squares estimator used in \cite{clyde;lee;2001} and \cite{taddy2016nonparametric}, as well as the instrumental variables estimator introduced in Section 2 of \cite{chamberlain2003nonparametric}.

Future work can extend our contribution. For example, it would be useful to determine which functionals are best estimated by the BLBB or SDBB (beyond empirical investigations), so that we can select and combine the methods as needed depending on the functionals we want to estimate. It would also be interesting to find strategies for determining when the sample size is big enough so that asymptotic methods (such as ANS, as defined in Section~\ref{simulations}) can be used to our advantage. Another interesting direction for further research would be designing data-subsetting strategies for datasets that have categorical variables with low observed frequency levels, which is an important practical issue, as we argue in Section~\ref{OPM2011}.

\section*{Supplementary Material}

The supplementary material has 6 sections: the first provides theoretical results for the processes proposed in Sections 2.1, 2.2, and 2.3; the second has a figure which details the Monte Carlo algorithm for performing lossless inference for the class of functionals described in Section 2.3; the third contains a scheme for lossless simulation for the example in Section 2.3 in \cite{chamberlain2003nonparametric}; the fourth part explains how to perform lossless inference for the Dirichlet-Multinomial process; the fifth includes a table with relative and absolute errors related to the linear regression coefficients estimated from the OPM-2011 dataset in Section 4.1; Finally, the sixth assesses the performance of the BLBB, SDBB, ANs, and AN approximating coefficients of a quantile regression fitted to the OPM-2011 dataset.

\section*{Acknowledgments}

We would like to thank Jerome P. Reiter, Lancelot F. James, and Peter M\"uller for helpful comments that improved the content and presentation of the article. We would also like to thank Alexander D. Bolton and John M. de Figueiredo for providing the OPM dataset and their valuable feedback. A. F. Barrientos was supported by grants from the National Science Foundation (ACI 1443014 and SES 1131897) and the Alfred P. Sloan Foundation (G-2-15-20166003).

\bibliographystyle{apalike}
\bibliography{ref}

\end{document}